\documentclass{jpsj3}

\title{Electron- and Hole-Doping Effects on $A$-site Ordered NdBaMn$_2$O$_6$}

\author{Yasuhiro \textsc{Miyauchi}, Mitsuru \textsc{Akaki}, $^{1}$Daisuke \textsc{Akahoshi}, \newline
and Hideki \textsc{Kuwahara}\thanks{E-mail: h-kuwaha@sophia.ac.jp}}

\inst{Department of Physics, Sophia University, Tokyo 102-8554
\\$^{1}$Department of Physics, Toho University, Funabashi 274-8510}

\abst{We have investigated electron- and hole-doping effects on $A$-site ordered perovskite manganite NdBaMn$_2$O$_6$, which has the $A$-type (layered) antiferromagnetic (AFM) ground state. 
Electrons (holes) are introduced by partial substitution of Ba$^{2+}$ (Nd$^{3+}$) with Nd$^{3+}$ (Ba$^{2+}$). 
Electron-doping generates ferromagnetic (FM) clusters in the $A$-type AFM matrix.
With increasing the electron-doping level, the volume fraction of the FM phase or the number of the FM clusters is abruptly increasing. 
In contrast, the $A$-type AFM phase is robust against the hole-doping, and no FM correlation is observed in the hole-doped NdBaMn$_2$O$_6$.}

\kword{$A$-site ordered perovskite, manganese oxide, hole doping, electron doping}

\begin{document}
\maketitle
\newpage

\section{Introduction}

Transition-metal oxides with perovskite structure $R_{1-x}Ae_xB$O$_3$ ($R$:\ rare-earth, $Ae$:\ alkaline-earth, and $B$:\ transition-metal) and their derivatives have been intensively studied in terms of technological applications as well as fundamental physics because of their rich electronic properties such as high-temperature superconductivity in cuprates and colossal magnetoresistance (CMR) effect in manganites \cite{1}.
The CMR is usually observed near the phase boundary between ferromagnetic (FM) metallic and charge/orbital-ordered insulating (COI) (or spin-glass insulating) states \cite{Tokura,E}. 
In Pr$_{0.55}$(Ca$_{1-y}$Sr$_y$)$_{0.45}$MnO$_3$, for example, the FM metallic and COI phases compete with each other in a bicritical manner around $y$ = 0.25\cite{CO-FM}.
Pr$_{0.55}$(Ca$_{1-y}$Sr$_y$)$_{0.45}$MnO$_3$ with $y$ = 0.2, whose ground state is the COI one near the phase boundary, undergoes a transition from the COI to FM metallic phases when external magnetic fields $H$ $\geq$ 30~kOe are applied \cite{CO-FM}.
The $H$-induced COI-FM metallic transition accompanies the huge reduction in the resistivity, which is referred to as the CMR.  
In conventional perovskite manganites, $R^{3+}$ and $Ae^{2+}$ occupy perovskite $A$-site at random. 
This structural randomness due to $A$-site solid solution considerably lowers the transition temperatures of the electronic phases far below room temperature (RT), which makes its practical application difficult. 
Therefore, in order to realize new electronic device using the CMR effect, such $A$-site randomness must be removed or considerably reduced.

$A$-site ordered perovskite $R$BaMn$_2$O$_6$ is a promising candidate for solving the problem. 
There exists no structural randomness in $R$BaMn$_2$O$_6$, since $R$O and BaO layers are regularly stacked along the $c$ axis (Fig.\ 1).
As a result, the electronic phase transition occurs above RT in $R$BaMn$_2$O$_6$ \cite{LB,Y-Sm2}. 
$R$BaMn$_2$O$_6$ ($R$ = Sm-Ho, and Y) undergoes the charge/orbital ordering transition at $T_{\rm CO}$ = 380-500~K.
In $R$BaMn$_2$O$_6$ ($R$ = Nd-La), the FM transition occurs at $T_{\rm C}$ = 300-350~K, and the $A$-type antiferromagnetic (AFM) phase appears as the ground state. 
The FM, COI, and $A$-type AFM phases coincide with each other in a multicritical manner at $R$ = Nd around RT \cite{Y-Sm2,NdBa1}. 
Y. Ueda group reported that $A$-site ordered Sm$_{1-x}$La$_{x+y}$Ba$_{1-y}$Mn$_2$O$_6$ and (NdBaSm)$_2$Mn$_2$O$_6$ exhibit the CMR at RT, the magnitude of which is larger than 1000~\% \cite{SmLaBa1,NdBaSm}. 
However, these compounds show the CMR in a high magnetic field of $H$ = 90~kOe, and the resistivity is gradually decreasing with increasing $H$, indicating that the CMR observed in these compounds does not arise from the $H$-induced COI-FM metallic transition. 
For technological applications using the CMR effect of $A$-site ordered $R$BaMn$_2$O$_6$, the $H$-induced COI-FM metallic transition, $i$.$e$.\ the CMR transition must be realized in much lower $H$ by use of the bicritical competition between the COI and FM metallic states. 
Therefor, it is important to develop the effects of eliminating only $A$-type AFM phase between the COI and FM metallic phases. 
In this study, we have elaborately investigated the electron- and hole-doping effects near the multicritical region in order to extinguish the $A$-type AFM phase of the $A$-site orederd  NdBaMn$_2$O$_6$.

\section{Experiment}

$A$-site ordered Nd$_{1+x}$Ba$_{1-x}$Mn$_2$O$_6$ ($-0.05$ \(\leq\) $x$ \(\leq\) 0.05) were prepared in a polycrystalline form by solid state reaction.
Here we refer to Nd$_{1+x}$Ba$_{1-x}$Mn$_2$O$_6$ with $x > 0$ and $x < 0$ nominally as "electron-" and "hole-" doped samples, respectively. 
The term "electron/hole-doping" used in this paper means that excess/deficient $e_{\rm g}$ electrons are introduced into the half-doped COI state in Nd$_{1+x}$Ba$_{1-x}$Mn$_2$O$_6$ ($x=0$, Mn$^{3.5+}$) by increasing/decreasing $x$ ($x>0$) from $x=0$, respectively.\cite{dop}
Mixed powders of Nd$_2$O$_3$, BaCO$_3$, and Mn$_3$O$_4$ with an appropriate molar ratio were first calcined at 1273~K for 12~h in Ar atmosphere.
After regrinding, the calcined powders were pressed into a pellet and sintered at 1573-1623~K for 12-24~h in Ar atmosphere to promote $A$-site cation ordering. 
The sintered pellet was then annealed at 973~K in O$_2$ atmosphere. 
Characterization of the synthesized samples was performed by powder X-ray diffraction (XRD) method at RT\@. 
The magnetic properties were measured using a Quantum Design, Physical Property Measurement System (PPMS).
The resistivity was measured using a standard four-probe method.

\section{Results and Discussion}

Figure 1 (a) shows the XRD patterns of $A$-site ordered Nd$_{1+x}$Ba$_{1-x}$Mn$_2$O$_6$ with $x$ = 0, 0.05 and $-$0.05 at RT\@.
In the XRD pattern of $x$ = 0, superlattice reflection is clearly seen at 2\(\theta\) = 11.5 degrees, which is strong evidence that Nd and Ba are regularly ordered.
Owing to Nd/Ba ordering, NdBaMn$_2$O$_6$ crystallizes in the tetragonal $P4/mmm$. 
Superlattice reflection due to $A$-site cation ordering is also observed in the XRD patterns of $x$ = 0.05 and $-$0.05\@. 
This indicates that both $x$ = 0.05 and $-$0.05 samples retain the $A$-site ordered perovskite structure in spite of partial (5 \%) substitution of Ba$^{2+}$ or Nd$^{3+}$ sites as shown in Fig.\ 1 (a).
We confirmed that all Nd$_{1+x}$Ba$_{1-x}$Mn$_2$O$_6$ ($-0.05 \leq x \leq 0.05$) synthesized in this study have the $A$-site ordered perovskite structure.
Lattice parameters of all samples remain almost unchanged as shown in Fig.\ 1~(b).

Let us begin with electron-doped NdBaMn$_2$O$_6$.
We show in Fig.\ 2 temperature dependence of magnetization (a) and resistivity (b) of Nd$_{1+x}$Ba$_{1-x}$Mn$_2$O$_6$ (0.0 \(\leq\) $x$ \(\leq\) 0.05). 
The cooling magnetization of the parent compound NdBaMn$_2$O$_6$ ($x$ = 0) abruptly increases with decreasing temperature toward 280~K and then shows a sudden decrease at 280~K\@.
The sudden decrease is due to the first-order $A$-type AFM transition \cite{Sm-Nd}, which accompanies large thermal hysteresis. 
The resistivity show a clear jump at the $A$-type AFM transition temperature $T_{\rm N}$. 
These results are consistent with the previous report \cite{NdBa1}. 
With increasing $x$ (electron-doping level), FM clusters are produced in the $A$-type AFM matrix.
The number of the FM clusters or the volume fraction of the FM phase is abruptly increasing below around 300~K with increasing $x$ as seen from Figs.\ 2 (a) and 3\@.  
The resistivity of $x$ = 0.05 does not show any clear jump indicative of the $A$-type AFM transition and drops by several orders of magnitude at low temperatures compared with that of $x$ = 0.
The magnetization of $x$ = 0.05 is fully saturated in $H$ = 80~kOe at 5~K, and its value is close to the expected value of 3.45~{$\mu_{\rm B}$} (Fig.\ 3). 
However, in $x$ = 0.05, the gradual magnetization reduction is observed under RT\@.
This result suggests that the $A$-type AFM phase does not totally vanish in $x$ = 0.05 and that the main FM phase coexists with the minor $A$-type AFM phase.
We note that the magnetic and transport properties of Nd$_{1+x}$Ba$_{1-x}$Mn$_2$O$_6$ with $0.05 < x \leq  0.12$ are not significantly different from those of $x$ = 0.05 (not shown).

We then show the magnetic and transport properties of hole-doped Nd$_{1+x}$Ba$_{1-x}$Mn$_2$O$_6$ ($-0.05$ \(\leq\) $x$ \(\leq\) 0.0) in Fig.\ 4\@.
The hole-doping effect is quite contrastive to the electron-doping effect. 
Hole-doping suppresses the FM correlation just above $T_{\rm N}$. 
On the other hand, the $A$-type AFM phase is robust against hole-doping. 
The magnetization reduction due to the $A$-type AFM order is clearly observed even for $x$ = 0.05, but $T_{\rm N}$ is reduced by about 10~K.  
As also clearly seen from the magnetization curves (Fig.\ 5), the FM component is not observed at all in hole-doped NdBaMn$_2$O$_6$. 
The resistivities of hole-doped NdBaMn$_2$O$_6$ show a clear jump at $T_{\rm N}$.

Here let us discuss the electron- and hole-doping effects of NdBaMn$_2$O$_6$.
It is experimentally confirmed that tetragonal lattice distortion expanding the $ab$-plane stabilizes the $x^2-y^2$ orbital ordered state with $A$-type AFM order \cite{x-y,kuwa}.
In the case of non-doped NdBaMn$_2$O$_6$, the $A$-type AFM state is stabilized by tetragonal lattice distortion that is enhanced by $A$-site cation ordering. 
Not only tetragonal lattice distortion but also the average Mn valence $V_{\rm Mn}$ also plays a significant role in the appearance of the $A$-type AFM state.
The $A$-type AFM phase is found in conventional ($A$-site disordered) perovskite manganites such as La$_{1-x}$Sr$_{x}$MnO$_3$ and Nd$_{1-x}$Sr$_{x}$MnO$_3$ in the over-doped region of $x \geq 0.5$\cite{LSMO,NSMO}, where $V_{\rm Mn}$ is larger than +3.5.
On the other hand, the $A$-type AFM phase totally disappears in the under-doped region of $x < 0.5$ ($V_{\rm Mn}$ $<$ +3.5).
$V_{\rm Mn}$ of electron-doped NdBaMn$_2$O$_6$ is smaller than +3.5($V_{\rm Mn}$ of non-doped NdBaMn$_2$O$_6$ is +3.5) so that the $A$-type AFM state is replaced by the FM one in electron-doped NdBaMn$_2$O$_6$. 
In hole-doped NdBaMn$_2$O$_6$, since $V_{\rm Mn}$ is larger than +3.5, the $A$-type AFM state persists in the hole-doped region.

In this study, we have demonstrated that the $A$-type AFM state that blocks the field-induced COI-FM metallic transition, $i.e$.\ the CMR effect, of the $A$-site ordered $R$BaMn$_2$O$_6$ can be easily destroyed by a slight amount of electron-doping. 
In fact, we have succeeded in achieving the CMR effect at RT for (Nd$_{0.7}$Sm$_{0.3}$)$_{1+x}$Ba$_{1-x}$Mn$_{2}$O$_{6}$($x=0.05$) although $H$ = 80~kOe is needed, which will be published elsewhere.
Further detailed study on the electron-doping effect on $R$BaMn$_2$O$_6$ near the multicritical region is needed to realize the RT CMR effect that is caused by much lower $H$ than that previously reported \cite{SmLaBa1,NdBaSm}.

\section{Conclusion}

In summary, we have investigated the electron(hole)-doping effect of $A$-site ordered NdBaMn$_2$O$_6$ by partial substitution of Ba$^{2+}$(Nd$^{3+}$)  with Nd$^{3+}$(Ba$^{2+}$). 
The $A$-type AFM phase is fragile against electron-doping, and electron-doping creates the FM clusters in the $A$-type AFM matrix.
With increasing the electron-doping level, the volume fraction of the FM phase or the number of the FM ciusters is abruptly increasing. 
On the other hand, no FM component is observed in hole-doped NdBaMn$_2$O$_6$, and the $A$-type AFM phase is robust against hole-doping.
Present results provide significant information for electronic phase control of $A$-site ordered perovskite manganites, which could lead to technological applications using the colossal magnetoresistance (CMR) effect.
\\

\section*{Acknowledgment}

This work was partly supported by Grant-in-Aid for JSPS Fellows, the Mazda Foundation, and the Asahi Glass Foundation.

\newpage
\begin{figure*}[tb]
\begin{center}
\includegraphics[width=0.4 \textwidth, clip]{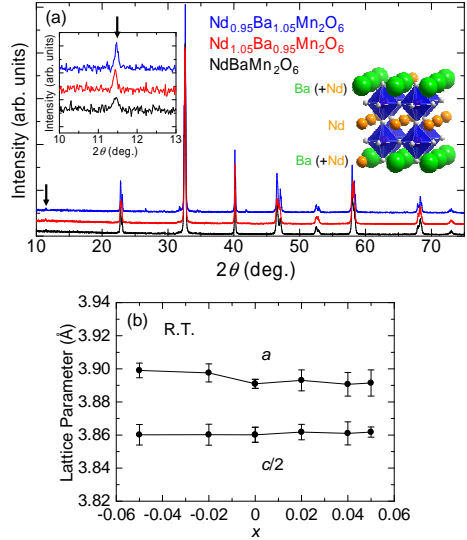}
\end{center}
\caption{(Color online) (a) X-ray-diffraction patterns of $A$-site ordered Nd$_{1+x}$Ba$_{1-x}$Mn$_2$O$_6$ with $x$ = 0, 0.05, and $-$0.05 at room temperature, and schematic structure of Nd$_{1.05}$Ba$_{0.95}$Mn$_2$O$_6$. 
The inset shows the enlargement of the diffraction profiles at 2\(\theta\) = 10-13 deg. Each pattern is vertically shifted for clarity. (b) Lattice parameters of $a$ and $c/2$ in the tetragonal $P4/mmm$ setting as a funnction of $x$ at room temperature.}
\label{fig1}
\end{figure*}

\begin{figure}[tb]
\begin{center}
\includegraphics[width=0.4 \textwidth, clip]{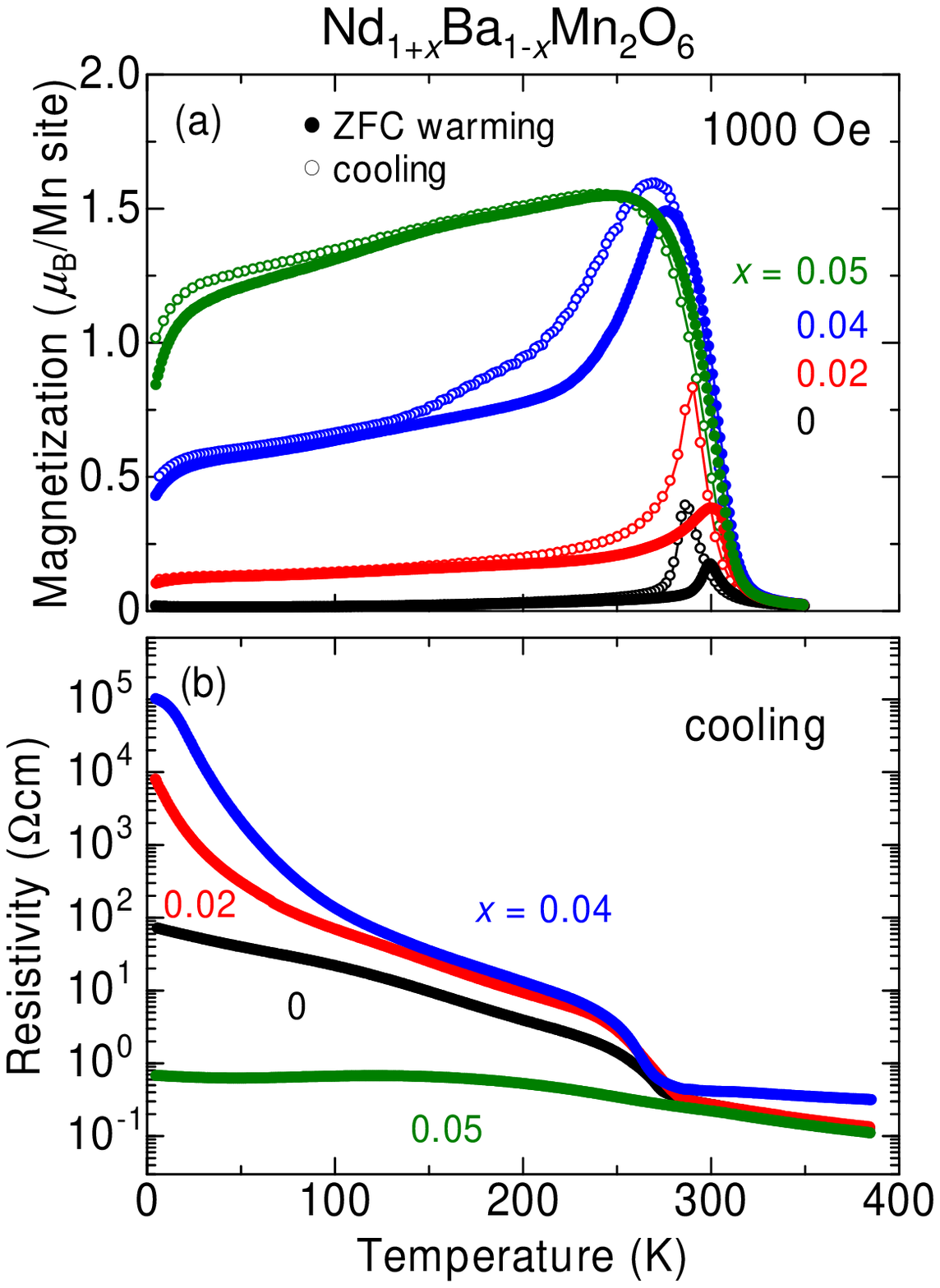}
\end{center}
\caption{(Color online) (a) Temperature dependence of the magnetization for Nd$_{1+x}$Ba$_{1-x}$Mn$_2$O$_6$ (0 \(\leq\) $x$ \(\leq\) 0.05) measured on cooling (open symbols) and on warming after zero field cooling (ZFC) (solid symbols).
(b) Temperature dependence of the resistivity for 0 \(\leq\) $x$ \(\leq\) 0.05 measured on cooling in a zero magnetic field.}
\label{fig2}
\end{figure}

\begin{figure}[tb]
\begin{center}
\includegraphics[width=0.4 \textwidth, clip]{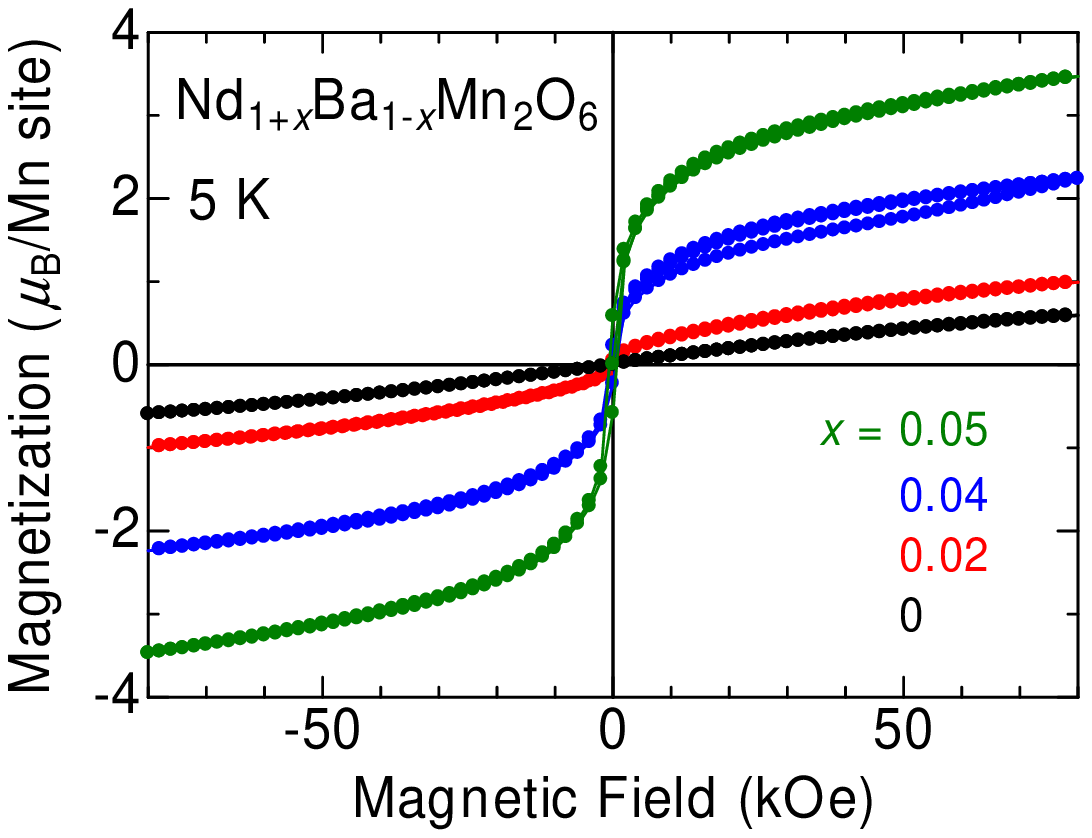}
\end{center}
\caption{(Color online) The magnetization curves of 
Nd$_{1+x}$Ba$_{1-x}$Mn$_2$O$_6$ (0 \(\leq\) $x$ \(\leq\) 0.05) at 5~K.\\}
\label{fig3}
\end{figure}

\begin{figure}[tb]
\begin{center}
\includegraphics[width=0.4 \textwidth, clip]{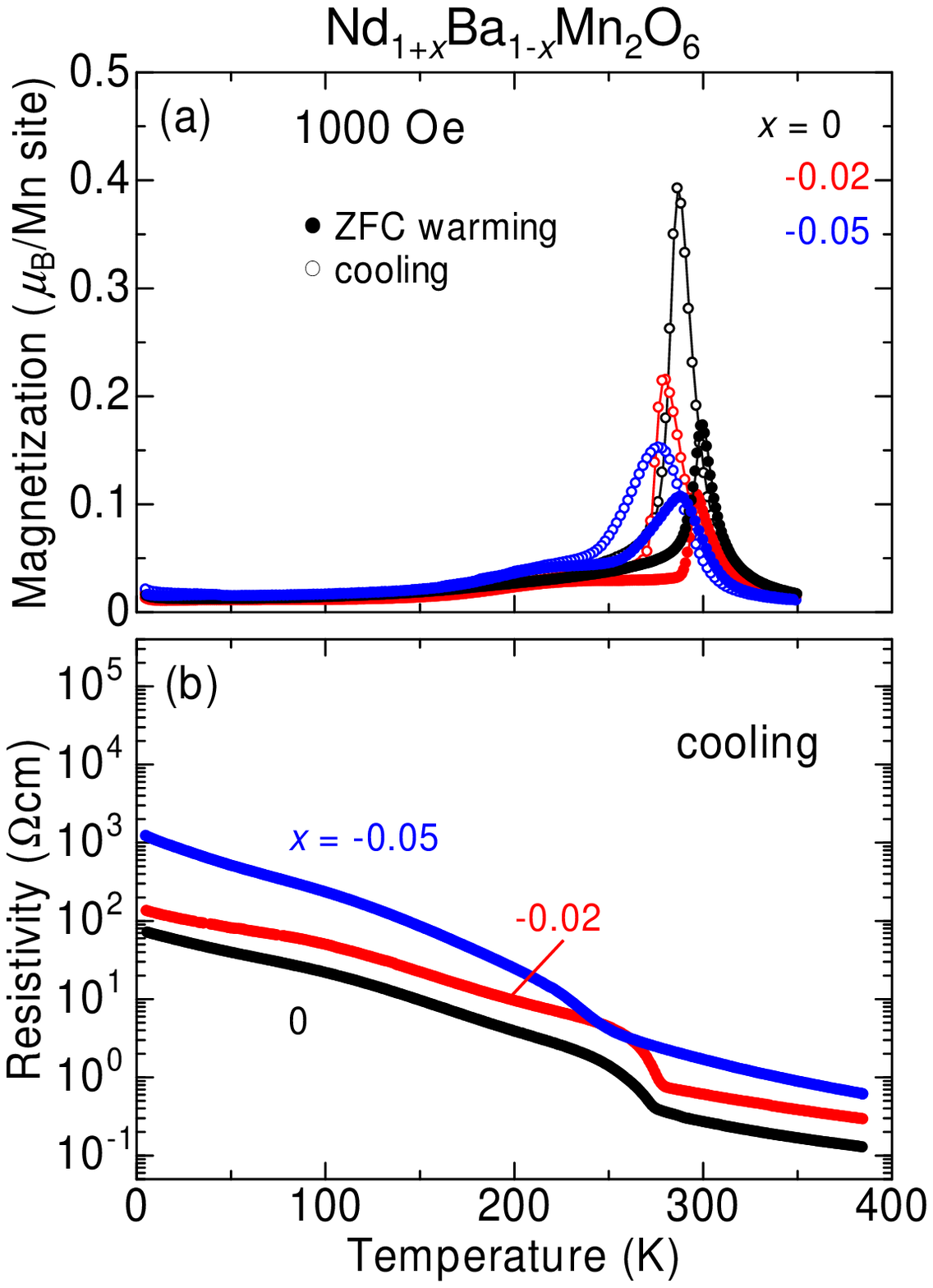}
\end{center}
\caption{(Color online) (a) Temperature dependence of the magnetization for Nd$_{1+x}$Ba$_{1-x}$Mn$_2$O$_6$ ($-$0.05 \(\leq\) $x$ \(\leq\) 0) measured on cooling (open symbols) and on warming after ZFC (solid symbols).
(b) Temperature dependence of the resistivity for $-$0.05 \(\leq\) $x$ \(\leq\) 0 measured on cooling in a zero magnetic field.}
\label{fig4}
\end{figure}

\begin{figure}[tb]
\begin{center}
\includegraphics[width=0.4 \textwidth, clip]{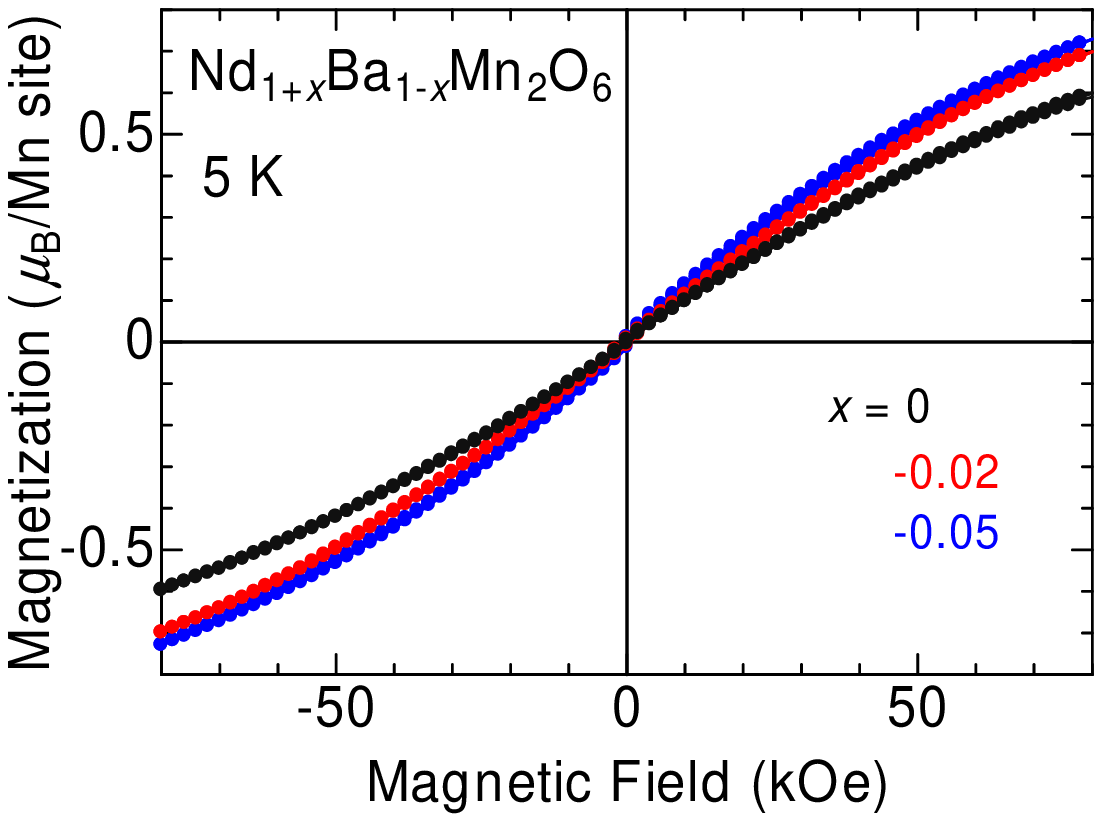}
\end{center}
\caption{(Color online) The magnetization curves of Nd$_{1+x}$Ba$_{1-x}$Mn$_2$O$_6$ ($-0.05$ \(\leq\) $x$ \(\leq\) 0) 
at 5~K\@.}
\label{fig5}
\end{figure}

\end{document}